\documentstyle[12pt]{article}

\newcommand{\leqsim}{\,\raisebox{-0.6ex}{$\buildrel < \over \sim$}\,}
\newcommand{\geqsim}{\,\raisebox{-0.6ex}{$\buildrel > \over \sim$}\,}
\newcommand{\con}{\rm constant}
\newcommand{\B}{\rm B}
\newcommand{\ba}{\begin{eqnarray}}
\newcommand{\be}{\begin{equation}}
\newcommand{{\df}}{\rm d}
\newcommand{\ea}{\end{eqnarray}}
\newcommand{\ee}{\end{equation}}
\newcommand{\eq}{\rm eq}
\newcommand{\F}{\rm F}
\newcommand{\ft}{{\scriptstyle (T)}}
\newcommand{\ftr}{{\scriptstyle (T_{\R})}}
\newcommand{\K}{\rm K}

\newcommand{\LR}{\stackrel{\rm L}{\rm R}}
\newcommand{\nhat}{\hat{n}}
\newcommand{\ox}{\otimes}
\newcommand{\pl}{\rm Pl}
\newcommand{\R}{\rm R}
\newcommand{\rad}{\rm rad}
\newcommand{\sB}{s\rm B}
\newcommand{\sF}{s\rm F}
\newcommand{\shat}{\hat{\sigma}}
\newcommand{\sK}{s\rm K}
\newcommand{\tev}{{\mbox{TeV}}}

\begin{document}
\thispagestyle{empty}
\begin{flushright}
{\tt DRAL-94-094\\ OUTP-94-20P\\ September 1994}
\end{flushright}
\vspace{5mm}
\begin{center}
{\LARGE Cosmological Constraints on\\
 \vspace{5mm} Perturbative Supersymmetry Breaking}\\ \vspace{15mm}
{\Large S.~A.~Abel$^{*}$\ and \ S.~Sarkar$^{\dag}$}\\ \vspace{10mm}
{\it $^{*}$Rutherford Appleton Laboratory \\Chilton, Didcot OX11 0QX, U.K.\\
\vspace{5mm}
$^{\dag}$Theoretical Physics, University of Oxford \\1 Keble Road,
            Oxford OX1 3NP, U.K.}
\end{center}
\vspace{2cm}
\begin{abstract}
\noindent
We discuss the cosmology of string models with perturbative
supersymmetry breaking at a scale of ${\cal O}$(TeV). Such models
exhibit Kaluza-Klein like spectra and contain unstable massive
gravitinos/gravitons. We find that considerations of primordial
nucleosynthesis constrain the maximum temperature following inflation
to be not much larger than the supersymmetry breaking scale. This
imposes conflicting requirements on the scalar field driving
inflation, making it rather difficult to construct a consistent
cosmological history for such models.
\end{abstract}
\vspace{4cm}
hep-ph/9409350
\newpage
\medskip \par
There has recently been much interest in the possibility of realistic
string theories with spontaneous {\it perturbative} supersymmetry breaking
\cite{FKP87,A90}. These theories are very predictive in that they
yield, in addition to the supersymmetric standard model, an entirely
new phenomenon with a striking experimental signature
\cite{FKP87}-\cite{AS94}. Specifically they contain a repeating
spectrum of Kaluza-Klein (KK) modes all the way up to the Planck
scale, whose spacing ($\epsilon \approx 1/2R$, where $R$ is the radius
of compactification) is comparable to the supersymmetry breaking scale
which is of ${\cal O}$(TeV). These modes can be excited at forthcoming
accelerators such as the LHC \cite{A93}, hence such models should be
of immediate interest to experimentalists. However there are open
questions concerning the cosmological viability of these models which
need to be addressed first. In this paper we will investigate whether
the Kaluza-Klein spectrum is consistent with cosmological constraints
on massive unstable relic particles \cite{EGS92}. (Note that the more
commonly discussed models with dynamical supersymmetry breaking in a
`hidden' sector are also constrained by similar cosmological
considerations \cite{DCQR93,RT94}.)
\medskip \par
The only previous relevant discussion on the cosmology of Kaluza-Klein
theories concentrated on the prospect that they may include absolutely
stable massive particles referred to as `pyrgons' \cite{KS94}. Such
particles reside on the first rung of the ladder of Kaluza-Klein
states, and are unable to decay because they carry a charge which is
not exhibited by any of the massless particles.  We shall not consider
such models since, as we shall see, `pyrgons' do not exist in string
theories with spontaneously broken supersymmetry. Although motivated
primarily by string theory, our discussion will apply to all
Kaluza-Klein theories in which the couplings are approximately
independent of winding number.
\medskip \par
For such a theory, the thermal history of the universe is radically
altered in the following way. We assume, as is usual, that there was
an inflationary DeSitter phase, followed by reheating to a temperature
$T_{\R}$ \cite{KT90}. In conventional supergravity, reheating results
in the production of gravitinos with number density proportional to
$T_{\R}$; the subsequent decays of the gravitinos can adversely affect
primordial nucleosynthesis and requiring that they not do so results
in an upper bound on the reheat temperature of $\sim 10^5$ TeV
\cite{W82}-\cite{ENS85}. After reheating, the entropy, which we shall
assume is subsequently conserved, is evenly spread out amongst the
strongly (as opposed to gravitationally) interacting KK modes and the
massless matter multiplets. At a temperature much higher than the KK
level-spacing ($T \gg \epsilon$), nearly all the entropy is in the KK
modes and almost none in the matter multiplets. Until the temperature
drops below the first KK level, the evolution of the universe is
therefore governed by the KK modes, whose contribution to the entropy
is continually decreasing as the temperature drops. During this period
there is production of massive gravitons and gravitinos which can only
decay to the massless (twisted) particles since their decays to
untwisted KK modes is kinematically suppressed. Under these
circumstances one might suspect that there is very severe bound on
$T_{\R}$ and this indeed turns out to be the case.
\medskip \par
Let us first present the `conventional' picture. Gravitinos are
generated at high temperatures by two-body scatterings and the
equation governing their number density is
\be
\label{evol1}
\dot{n}_{3/2} + 3\,H\,n_{3/2} = \langle \sigma v \rangle n_{\rad}^2 -
 \frac{n_{3/2}}{\tau_{3/2}}\ ,
\ee
where $H$ is the Hubble expansion rate and $\langle \sigma v
\rangle$ is the thermally averaged cross-section for
gravitino production in the radiation bath of number density
$n_{\rad}$. The gravitino lifetime $\tau_{3/2}$ is given at rest by
\cite{CR91}
\be
\tau_{3/2} \sim M_{\pl}^2/m_{3/2}^3 \simeq 10^5\ \sec\
 \left(\frac{m_{3/2}}{\tev}\right)^{-3} ,
\ee
where $M_{\pl} \equiv G_{\rm N}^{-1/2} \simeq 1.22 \times 10^{16}$
TeV. Given the effective $N=1$ supergravity couplings,
\ba
\label{prod}
\delta L&=&\frac{\sqrt{2 \pi}}{2 M_{\pl}}{\overline\lambda}_a \gamma^\rho
 \sigma^{\mu\nu} \psi_\rho F_{\mu\nu}^a  \mbox{ + h.c.}  \nonumber\\
&&+ \frac{\sqrt{2 \pi}}{ M_{\pl}}{\overline\psi}_\rho \gamma^\mu \partial_\mu
 z^i \gamma^\rho\psi_i \mbox{ + h.c.}\ ,
\ea
it can be shown that $\langle \sigma v \rangle \sim (8
\pi/M_{\pl}^2)$ at temperatures $T \ll M_{\pl}$ \cite{EKN84}. The
radiation density is given by
\be
n_{\rad}=g\ft\ \frac{\zeta{\scriptstyle(3)}T^3}{\pi^2} ,
\ee
where $g\ft$ counts the relativistic degrees of freedom contributing
to the total number density and is constant above temperatures of
${\cal O}$(TeV) for the minimal supersymmetric standard model (MSSM), with
\be
g\ftr = \hat{g}= 427/2\ .
\ee
Assuming
the canonical radiation-dominated evolution at this time, we can
solve eq.(\ref{evol1}) to obtain
\be
Y_{3/2}\ft \equiv \frac{n_{3/2}}{n_{\rad}} =
 \frac{g_s\ft}{g_s\ftr}\ \frac{n_{\rad}\ftr \langle\sigma v\rangle}{H\ftr}\
 \exp (-t/\tau_{3/2})\ ,
\ee
where time is related to temperature as
\be
t = 2.42 \times 10^{-12}\ [g_s\ft]^{-1/2}\ \sec\
\left(\frac{T}{\tev}\right)^{-2} ,
\ee
and the factor $g_s\ft /g_s\ftr$ takes into account the decrease in the
number of relativistic degrees of freedom, given constant total entropy
\be
s R^3 = g_s\ft \frac{\pi^2 T^3}{30} R^3 = \con\ .
\ee
Note that we have taken $g_s\ft$ to also be the number of degrees
of freedom determining the total energy density, as is appropriate at
temperatures above a few MeV (when the neutrinos decouple). For the
MSSM, one has
\be
g_s\ftr = \hat{g}_s = 915/4\ ,
\ee
at high temperatures when all particles are relativistic.
\medskip \par
Now let us consider the cosmological evolution when KK modes are
present. Above the supersymmetry breaking scale the number of
relativistic degrees of freedom is now no longer constant. The KK
modes are labelled by quantum numbers of internal momenta/charges
which are of the form
\be
P_{\LR}=\frac{n}{R}\pm \frac{mR}{2}\ ,
\ee
where $R$ represents some internal radius of compactification. The
winding modes ($m\neq 0$) have masses of ${\cal O}(M_{\pl})$ and need
not be considered further, while the particles in the $n$th KK mode
have masses $m_n \sim n \epsilon$. Roughly speaking, whenever the
temperature is raised by $\epsilon$, two new levels of (gauge
interacting) KK excitations becomes relativistic, so that the number
of degrees of freedom increases {\it linearly} with temperature. We
can allow for this by writing
\be
\label{KKdof}
g\ft = (\hat{g}+ \frac{T}{\epsilon} g_{\K})\ , \hspace{1cm}
g_s\ft = (\hat{g}_s+ \frac{T}{\epsilon} g_{\sK})\ .
\ee
The constants $g_{\K}$, $g_{\sK}$ are determined by evaluating the
number density and entropy density, respectively, of the plasma. For
example consider the number density of KK modes in equilibrium:
\be
n_{\eq}\ft = \sum_{n} \frac{g_i}{(2 \pi)^3} \int {\df}^3 p\
 \frac{1}{\exp(\sqrt{p^2 + n^2\epsilon^2}/T) \pm 1}\ ,
\ee
where $g_{i}$ is the total number of interacting degrees of freedom in
any KK level. Using various redefinitions, this becomes
\be
n_{\eq}\ft \sim \frac{g_i T^4}{\pi^2 \epsilon} \int_0^{\infty}
 {\df} x \int_{\epsilon/T}^{\infty}
 {\df} y\ \frac{x^2}{\exp{\sqrt{x^2 + y^2}} \pm 1}\ ,
\ee
where we have approximated the sum at small $\epsilon /T$ by an
integral, and included a factor of two for positive and negative
values of internal momentum (defined above as $n$). The integral
becomes temperature independent for $T\gg \epsilon $, and we find that
it deviates by less than 5\% from the $T^4$ behaviour for high values
of $T/\epsilon\ (\gg 2)$. The limiting values (when $\epsilon /T
\rightarrow 0$) may be determined analytically to be
\be
g_{\K}= g_{i} (\chi_{\F} + \chi_{\B})\ ,
\ee
where
\be
\chi_{\F} = 7 \pi^5 / 480\,\zeta (3) = 3.71\ , \hspace{1cm}
\chi_{\B} = \pi^5 / 60\,\zeta (3) = 4.24\ .
\ee
In a similar fashion, the contribution of the KK modes to the total
entropy may be used to determine $g_{\sK}$. Using $s \equiv (\rho +
p)/T$, we find that
\be
s_{\eq}\ft \approx \frac{g_i T^4}{\pi^2 \epsilon} \int_0^{\infty}
 {\df} x \int_{\epsilon/T}^{\infty}
 {\df} y\ \frac{x^2}{\exp\sqrt{x^2 + y^2} \pm 1}\
 \frac{(4x^2/3 + 3 y^2)}{\sqrt{x^2 + y^2}}
\ee
which gives limiting values of
\be
g_{\sK}= g_i (\chi_{\sF} + \chi_{\sB})\ ,
\ee
where
\be
\chi_{\F} = 10125\,\zeta (5) / 64 \pi^3 = 5.29\ , \hspace{1cm}
\chi_{\B} = 675\,\zeta (5) / 4 \pi^3 = 5.64\ .
\ee
In the spontaneously broken string theories, each KK level comes in
$N=4$ multiplets, so that KK gauge bosons contribute 8 bosonic and 8
fermionic degrees of freedom in the vector and fermionic
representations of SO(8) respectively. In the minimal case in which
the KK excitations are in SU(3) $\ox$ SU(3)$_{\rm c}$ multiplets
\cite{AB94}, this gives
\be
g_{\K} = 128 (\chi_{\F} + \chi_{\B}) = 1018, \hspace{1cm}
g_{\sK} = 128 (\chi_{sF} + \chi_{sB}) = 1400\ .
\ee
Of course there are additional contributions from higgs multiplets
which are also expected to have KK excitations, so we shall consider
these values to be a lower limit. To find the time-temperature
relation, we assume that after inflation the metric is of the usual FRW
form, with all the relativistic degrees of freedom in chemical
equilibrium and therefore at the same temperature. Entropy
conservation then gives
\be
s R^3 = \frac{g_{\sK}\ft}{\epsilon} \frac{\pi^2 T^4}{30} R^3 = \con,
\ee
and in particular
\be
\label{hrel}
H = - \frac{4}{3} \frac{\dot{T}}{T}\ .
\ee
Thus the Hubble parameter is
\be
H\ft = 1.66\ \sqrt{\frac{g_{\sK}}{\epsilon}}\ \frac{T^{5/2}}{M_{\pl}}\ .
\ee
Differentiating with respect to time and substituting eq.(\ref{hrel}), we
find
\be
t\ft = \frac{8}{15 H\ft}\ .
\ee
For the minimal value of $g_{\sK}$ above, this becomes
\be
\label{time-temp}
t = 6.9 \times 10^{-14}\ \sec \left(\frac{\epsilon}{\tev} \right)^{1/2}
    \left(\frac{T}{\tev}\right)^{-5/2} ,
\ee
at temperatures $T\gg \epsilon$.
\medskip \par
In order to ascertain the abundance of massive gravitons/gravitinos,
we need to identify the processes which can contribute to their
manufacture and decay. Vertices between KK modes (which come from
untwisted sectors of the string theory) must satisfy the condition
that their internal momenta/charges are conserved, and are simply
proportional to the string coupling constant. In addition vertices can
exist between untwisted modes and twisted (massless) matter
multiplets, with couplings $g_n \propto g \delta^{-m_n^2/M_{\pl}^2}
\approx g$, where $\delta$ is some constant depending on the type of
compactification \cite{HV87}. For masses much less than $M_{\pl}$
these are clearly unsuppressed, so that we can write down effective
terms for the coupling of the KK modes to each other, and to the
massless multiplets. The creation of KK gravitons and gravitinos goes
via effective four particle interactions. For example the cross
section for
\be
A^{a}_{k} + A^{b}_{l} \rightarrow \lambda^{c}_{m} + \psi^{\mu}_{n}
\ee
goes as
\be
\sigma \sim \frac{g^2}{64 \pi} f_{abc} f^{abc} \frac{8 \pi}{M_{\pl}^2}
 \delta (n+m-k-l)\ ,
\ee
where the integers $k, l, m, n$ label KK modes, and we have omitted
numerical factors of ${\cal O}(1)$ coming from the trace over
solutions to the Rarita-Schwinger equation and phase-space
integrations. We therefore write the total cross section for
$n$-gravitino production from $k$ plus $l$ interactions as
\be
\label{sigma}
\sigma_{klmn} = \shat\ \delta (n+m-k-l)\ .
\ee
where $\shat$ is a factor of ${\cal O}(8\pi/M_{\pl}^2)$, which is
dependent on the details of the model, but independent of the
KK-modes. In addition there is a contribution coming from the massless
sector which we shall neglect since there is only one such sector.
\medskip \par
The massive $n$-gravitinos/gravitons may decay into either two
massless twisted states, or to untwisted $l$ plus $n-l$-states. In the
first case the decay rate is found to be,
\be
\Gamma_{\rm twisted} \approx \left(\frac{m_n^3}{M_{\pl}^2}\right).
\ee
The decay to untwisted states is kinematically suppressed however.
Consider a positive-$n$ state decaying to an $l$-state plus an
$n-l$-state.  If $l$ is negative then the masses of the products is
$|n|\epsilon +2|l|\epsilon $ which is larger than the mass,
$|n|\epsilon$, of the decaying particle. If $l$ is also positive, then
sum of the product masses is equal to that of the decaying particle.
The decay rate is proportional to the momentum of the decay products
in the centre of mass frame:
\be
\Gamma_{\rm untwisted} \propto |p| = \sqrt{1+\frac{[l^2-(n-l)^2]^2}{n^4}
 - 2 \frac{[l^2 + (n-l)^2]}{n^2}}\ \frac{\epsilon |n|}{2} = 0\ .
\ee
As discussed in ref.\cite{AS94}, there may be a mass splitting of
${\cal O}(\epsilon)$ in any KK level, so that the decay rate to
untwisted modes will generally be suppressed by a factor $1/|n|$. When
the sum over $l$ is taken this decay mode may be of the same order as
the decay to twisted states. We shall therefore take the lifetime of
$n$-gravitons/gravitinos to be
\be
\tau_n \sim 9.8 \times 10^4\ \sec \left(\frac{\epsilon}{\tev}\right)^{-3}
 |n|^{-3}\ ,
\ee
corresponding to the lifetime for decay of standard gravitinos into
photons and photinos \cite{CR91}. Inclusion of all the strongly
interacting, twisted, final states would speed up the decays by a
small factor. For example, if the twisted products consisted of all
the matter and Higgs particles in the MSSM, then the above would be
reduced by a factor $12/49$.
\medskip \par
With these estimates we are ready to tackle the evolution of the
$n$-graviton/gravitino number density $n_n$. This is governed by the
equation \cite{B88}
\ba
\label{evol0}
\dot{n}_n+ 3 H n_n +
 \frac{n_n}{\tau_n}
&=& \sum_{mkl}  \sum_{spins}
\int\int\int\int
 \frac{\df^3 q_k}{(2\pi)^3}\frac{\df^3 q_l}{(2\pi)^3}
 \frac{\df^3 q_m}{(2\pi)^3}\frac{\df^3 q_n}{(2\pi)^3}
 (2\pi)^4 \delta^4 (q_k+q_l-q_m-q_n)
\nonumber\\
& & f_k f_l (1 - f_m) \left| M_{kl\rightarrow mn} \right|^2
\ea
where the $f_i$ are the occupation numbers and we have assumed that
the massive particles are non-relativistic when they decay. This
equation describes the creation of a gravitationally interacting
$n$-state plus a gauge interacting $m$-state, from gauge interacting
$k$ plus $l$-states. We have taken the occupation number of the
$n$-state to be negligible (since it is never in equilibrium), so that
the reverse process does not occur. We may reasonably adopt
equilibrium distributions for the three remaining ($k,l,m$) states,
omit the Pauli blocking factor for the $m$-state, and rewrite
eq.(\ref{evol0}) as
\be
\label{evol2}
\dot{n}_n+ 3 H n_n  +
 \frac{n_n}{\tau_n}=  \sum_{mkl}\sigma_{klmn}\,
 n_{\eq}{\scriptstyle (k,T)}\ n_{\eq}{\scriptstyle (l,T)},
\ee
where, since we are concerned with the longest lived states, we have
taken $v \approx c$. Although approximate, this expression has the
correct temperature dependence and we shall bury our approximations in
the parameter $\shat$ which is still of ${\cal O}(8\pi/M_{\pl}^2)$.
Defining
\be
n_n = \nhat_n\ \exp (-t/\tau_n),
\ee
and using eq.(\ref{hrel}), we find
\be
\label{evol3}
T^4 \frac{{\df}}{{\df} t}\left(\frac{\nhat_n}{T^4}\right)=
 \sum_{mkl} \sigma_{klmn}\ n_{\eq} {\scriptstyle (k,T)}\
 n_{\eq} {\scriptstyle (l,T)}\ \exp (-t/\tau_n)\ .
\ee
The number density of the individual $k$ and $l$ states follow some
equilibrium curve which obeys, e.g.
\be
n_{\rad} \approx n_{\eq}\ft
 = \sum_{k=-\infty}^{+\infty} n_{\eq}{\scriptstyle (k,T)}\ .
\ee
Since the epoch of nucleosynthesis is much later than the time at
which these particles are created, we take the exponential factor to
be unity for the purposes of calculating the initial abundance.
Performing the summations in eq.(\ref{evol3}) and integrating from
$T_{\R}$ to $T = \epsilon$ gives
\be
Y_{n}\ft=\frac{8}{9}\ \frac{g_s\ft\,g_{\K}}{{\hat g}_s\,{\hat g}}\
 \frac{\shat\,n_{\rad}\ftr}{H\ftr}\ \exp (-t/\tau_n) ,
\ee
where, again, there is a factor to account for the change in photon
number density between $\epsilon$ and the final temperature
$T<\epsilon$.  Note that the final density of $n$-gravitons/gravitinos
is proportional to $T_{\R}^{3/2}$. For typical parameter values
($\shat = 8 \pi/M_{\pl}^2 $, $g_{\K} = 1018$, $g_{\sK} =
1400$), this is
\be
Y_{n}\ft \sim 3 \times 10^{-16}\
 \left(\frac{\epsilon}{\tev} \right)^{1/2}
 \left(\frac{T_{\R}}{\tev} \right)^{3/2}\ \exp (-t/\tau_n)
\ee
at $T \leqsim \epsilon$; the relic abundance during nucleosynthesis
would be smaller by a factor ${g_s\ft}/g_s{\scriptstyle (T_0)}$ where
${g_s {\scriptstyle (T_0)}}= 43/11$ is the effective number of
entropic degrees of freedom at $T \ll m_{e}$ (taking into account the
three decoupled relativistic neutrino species). This is close to the
value of 3.36 for the effective number of degrees of freedom
contributing to the total energy density, so for convenience we have
ignored the small difference.
\medskip \par
The energy density in the relic KK modes decreases as $T^{17/3}$ for
$T > \epsilon$, i.e. {\em faster} than the energy density in
`radiation' (including KK excitations) which goes as $T^5$. (At $T <
\epsilon$ the former decreases as $T^{16/3}$, while the latter does so
as $T^4$.) Therefore the universe will become `radiation' dominated
when
\be
\sum_{n} |n|\ \epsilon\ Y_n\ft\ n_{\rad}\ft < g_s\ft\ \frac{\pi^2 T^4}{30}\ .
\ee
where $g_s\ft$ is taken from eq.(\ref{KKdof}). Anticipating the bounds
which we will find on $T_{\R}$, the temperature at equality will be
higher than $\epsilon$, therefore the appropriate time-temperature
relationship is eq.(\ref{time-temp}). Since at late times, the number
density of KK modes is dominated by the lighter (slowly decaying)
particles which have a small exponential suppression factor in $Y_n$,
we may approximate the sum on the left above by an integral,
\ba
\sum_{n=-\infty}^{\infty} |n|\ \exp (-t n^3/\tau_1)
&\approx & \int_{0}^{\infty} {\df}^2 x\ x\ \exp (-x^3)
 \left(\frac{t}{\tau_1}\right)^{-2/3}\nonumber\\
 &= &\frac{2}{3}\ \Gamma\left(\frac{2}{3}\right)
    \left(\frac{t}{\tau_1}\right)^{-2/3} .
\ea
This expression is valid for $t \leqsim \tau_1$; at later times, the
KK modes have all decayed, i.e. their number is exponentially
suppressed. Therefore `radiation' domination will occur at
\be
T \geqsim 1.1 \times 10^{6}\ \tev \left(\frac{\epsilon}{\tev}\right)^{5/2}
\left(\frac{T_{\R}}{\tev}\right)^{-1}
\ee
corresponding to a time
\be
t \leqsim 5.3 \times 10^{-29}\ \sec
\left(\frac{\epsilon}{\tev}\right)^{1/2}
\left(\frac{T_{\R}}{\tev}\right)^{5/2}\ .
\ee
It may appear surprising that the higher the reheat temperature, the
{\it less} likely the gravitinos are to matter dominate at a late
time. This can be explained as follows; there is only a limited amount
of entropy, and when $T_{\R}$ is high, more of it is initially
distributed in heavier modes with higher KK number. Since these modes
decay more rapidly, they are able to matter dominate only at very
early times.
\medskip \par
We can now examine the effect of the decaying particles on the
abundances of the light elements. The effect of hadronic decays
occuring at the beginning of the nucleosynthesis era on neutron-proton
interconversions has been studied in detail in ref.\cite{RS88}. In the
time-interval $1 \leqsim t \leqsim 10^2 \sec$, the requirement that
the $^{4}$He mass fraction not be increased above $25 \%$ translates
into the requirement (see figure 3 in ref.\cite{EGS92}):
\be
\sum_{n} |n|\ \left(\frac{\epsilon}{\tev}\right)\ Y_n\ft\ n_{\rad}\ft
 \leqsim 2 \times 10^{-11},
\ee
which corresponds to the bound
\be
T_{\R} \leqsim 21\ \tev \left(\frac{\epsilon}{\tev}\right)^{1/3} .
\ee
For $t \geqsim 10^4 \sec$, the photodissociation of $^{2}$H due to the
radiation cascades triggered by the decaying particles impose the
constraint \cite{EGS92}
\be
\sum_{n} |n|\ \frac{\epsilon}{\tev}\ Y_n\ft\ n_{\rad}\ft \leqsim
10^{-13},
\ee
or
\be
T_{\R} \leqsim 2.7\ \tev \left( \frac{\epsilon}{\tev} \right)^{1/3}\ .
\ee
One should however bear in mind that the approximation we used to
calculate the KK number density breaks down when $T_{\R}$ becomes
comparable to $\epsilon$ since the number density is then suppressed
by a Boltzmann factor. We can however justifiably assert that the
maximum temperature which the universe reached cannot have
significantly exceeded the supersymmetry breaking scale.
\par
A precise formulation of the early history of the universe is lacking
for these models, however it is clearly essential that there be an
inflationary phase \cite{KT90}, followed, as we have shown, by
reheating to a temperature no greater than the supersymmetry breaking
scale. We now argue that this imposes a conflicting set of
requirements on the scalar field $\Phi$ which is presumed to drive
inflation. In order to account for the amplitude of the
scale-invariant density fluctuations probed by {\sl COBE} \cite{W92},
$m_{\Phi}$ is required to be of ${\cal O}(10^8)$ TeV \cite{O90}. For
the reheat temperature to be low, the inflaton must then be very
weakly coupled to matter fields. However even assuming only
gravitational couplings, i.e. a decay rate $\Gamma_{\Phi}\ \sim
m_{\Phi}^3/M_{\pl}^2$, the reheat temperature cannot be reduced below
\be
  T_{\R} \sim \left[\frac{g_{\sK}\ft}{\epsilon}\right]^{-1/4}\
                (\Gamma_{\Phi} M_{\pl})^{1/2}
   \sim 10^3\ \tev\ \left(\frac{\epsilon}{\tev}\right)^{1/4}\
         \left(\frac{m_{\Phi}}{10^8 \tev}\right)^{3/2} ,
\ee
where $T_{\R}$ has been obtained by equating $\Gamma_{\Phi}$ to the
Hubble rate $H$. In fact, the only mass scale in these models, apart
from the Planck scale, is the supersymmetry breaking scale. One could
then envisage a scenario with two epochs of inflation \cite{RT94}; the
first stage to create the correct level of density fluctuations, and
the second, with $m_{\Phi} \sim \epsilon$, to remove the KK
states. However assuming gravitational couplings as above, the reheat
temperature is then of ${\cal O}$(keV), i.e. too low for even
primordial nucleosynthesis to occur.  On the other hand if the
inflaton is coupled directly, the reheat temperature would be too
high, unless the gauge coupling is unnaturally small, of ${\cal O}
(10^{-4}$) \cite{RT94}. A way out may be to introduce an intermediate
scale of unification, with a low-mass `flaton' singlet remaining after
symmetry breaking \cite{EEE89}. However the reheat temperature is then
of ${\cal O}$(MeV), i.e. barely high enough for nucleosynthesis, and
further, the Affleck-Dine mechanism must be invoked in order for the
baryons to be synthesized just beforehand.
\par
In conclusion, it appears difficult to construct a consistent
cosmological history for models with spontaneous supersymmetry
breaking. In view of their many other advantageous features
\cite{A90}, possible resolutions to this problem should be pursued.

\vspace{1cm}
\noindent
{\bf\Large Acknowledgement} \hspace{0.3cm} One of us (SAA) would like
to thank M.~Quir\'os for useful discussions, and the Instituto de
Estructura de la Materia, Madrid for hospitality whilst part of this
work was completed. SS was supported by a SERC Advanced Research
Fellowship.

\end{document}